# Synergistic Effect of One- and Two- dimensional Connected Coral-like $Li_{6.25}Al_{0.25}La_3Zr_2O_{12}$ in PEO-Based Composite Solid State Electrolyte

Jun Cheng,[a] Guangmei Hou,[a] Qing Sun,[a] Xiaoyan Xu,[a] Linna Dai,[a] Jianguang Guo,[a] Zhen Liang,[a] Deping Li,[b] Jianwei Li,[a] Xiangkun Nie,[a] HuanHuan Guo,[a] Zhen Zeng[a], Xueyi Lu[c],* and Lijie Ci[a,b],*

[a] SDU & Rice Joint Center for Carbon Nanomaterials, Key Laboratory for Liquid-Solid Structural Evolution & Processing of Materials (Ministry of Education), School of Materials Science and Engineering, Shandong University, Jinan 250061, China

[b] School of Materials Science and Engineering, Harbin Institute of Technology (Shenzhen), Shenzhen, 518055, China

[c] International Center for Materials Nanoarchitectonics, National Institute for Materials Science Namiki 1-1, Tsukuba 305-0044, Japan

Email: lu.xueyi@hotmail.com (X.L.)
Email: lci@sdu.edu.cn (L.C.)

# Abstract

As one of the most promising next-generation energy storage device, lithium metal batteries have been extensively investigated. However, the poor safety issue and undesired lithium dendrites growth hinder the development of lithium metal batteries. The application of solid state electrolytes has attracted increasing attention as they can solve the safety issue and partly inhibit the growth of lithium dendrites. Polyethylene oxide (PEO)-based electrolytes are very promising due to their enhanced safety and excellent flexibility. However, PEO-based electrolytes suffer from low ionic conductivity at room temperature and can't effectively inhibit lithium dendrites at high temperature due to the intrinsic semi-crystalline properties and poor mechanical strength. In this work, a novel coral-like $Li_{6.25}Al_{0.25}La_3Zr_2O_{12}$ (LALZO) is synthesized to use as an active ceramic filler in PEO. The PEO with LALZO coral (PLC) exhibits increased ionic conductivity and mechanical strength, which leads to the uniform deposition/stripping of lithium metal. The Li symmetric cells with PLC cycle for 1500 h without short circuit at 50 °C. The assembled LiFePO4/PLC/Li batteries display excellent cycling stability at both 60 °C and 50 °C. This work reveals that the electrochemical properties of organic and inorganic composite electrolyte can be effectively improved by tuning the microstructure of the filler, such as the coral-like LALZO architecture.

# Introduction

Lithium (Li) metal has been extensively considered as the most promising anode of lithium battery due to its high theoretical specific capacity (3860 mA h g$^{-1}$) and the lowest electrode potential (~3.040 V vs. the standard hydrogen electrode), and as one of the most promising next-generation rechargeable battery, lithium metal batteries (LMBs) have been studied for several decades.[1] However, there are many barriers hinder the practical application of lithium metal batteries. In traditional lithium metal batteries with liquid organic electrolytes, the side reactions and uneven plating/stripping of lithium during cycles cause undesired Li dendritic growth or mossy Li generation, which can lead to internal short circuiting or even explosion.[2,3]

The application of solid state electrolytes in LMBs has attracted increasing attention as they can solve the safety issue and effectively prevent the formation of unstable solid state electrolyte interphase (SEI).[4] In addition to inorganic ceramic solid electrolytes, solid polymer electrolytes, such as polyethylene oxide (PEO), are becoming increasing promising due to their excellent flexibility.[5] However, due to the intrinsic semi-crystalline properties and poor mechanical strength, PEO electrolyte suffers low room-temperature ionic conductivity and insufficient capability of inhibiting lithium dendrites. Therefore, a great deal of approaches have been adopted to overcome this issues.[6-10] Compositing ceramic fillers within PEO matrix is an effective approach to improve the ionic conductivity and mechanical properties.[8,10-12] The existence of fillers leads to the decrease of crystallinity of PEO, which further improve the ionic conductivity. Ceramic fillers can be divided into two categories：inactive ceramic filler and active ceramic fillers which exhibit ion conduction property. Adding inactive ceramic fillers, such as $Al_2O_3$,[13] $ZrO_2$,[14] and $SiO_2$[12] to establish solid composite electrolytes (CPE) are valid strategies. However, low ionic conductivities are still observed in these composite electrolytes with inactive ceramic fillers. Compare to inactive ceramic fillers, adding inorganic solid state electrolytes into the PEO matrix can be more effective to further improve the ionic conductivity since they can provide additional fast paths for lithium ion transmission. In previous studies, $Li_7La_3Zr_2O_{12}$ (LLZO)[15], $Li_{1.5}Al_{0.5}Ge_{1.5}(PO_4)_3$ (LAGP)[16] and $Li_{1+x}Al_xTi_{2-x}(PO_4)_3$ (LATP)[17] have been demonstrated not only useful to improve ionic conductivity of PEO electrolyte, but also increase its strength. And the composite electrolyte can effectively suppress the growth of undesired lithium dendrite. Besides, The interfacial resistance between electrolyte and lithium metal anode can be effectively reduced by compositing active fillers.[18] For instance, Chen et al. reported a composite electrolyte with LLZTO particles as the filler and found it is favorable to suppress the growth of Li dendrites.[18] Zhao et al. chose four types of LAGP with different particle sizes as active fillers to fabricate PEO/LAGP composite electrolytes. The results indicate that adding LAGP particles provides a positive effect to the ionic conductivity of composite electrolyte.

Besides ceramic particles, the ceramic nanowires and ceramic three-dimensional framework have attracted interests of many researchers due to its obviously elevation of the ionic conductivity. He et al. applied PEO as matrix of composite electrolyte embedded with LLZO nanowires, and verified that LLZO nanowires in PEO matrix can enhance the ionic conductivity and mechanical strength of composite electrolyte efficiently. And the addition of LLZO nanowires can also contribute to uniform deposition of lithium during cycle. Hu et al. reported composite electrolytes with a 3D ceramic textiles derived from templates, exhibiting the advantages of 3D lithium conductive ceramics in composite polymer electrolytes.

Ceramic nanoparticles[19], nanowires[17,20,21], nanosheets[22], and three-dimensional frameworks[23-25] as the filler in PEO have been widely studied, and their positive effects to the ionic conductivity and mechanical properties of composite electrolyte have been demonstrated. Herein, we report the synthesis of a coral-like $Li_{6.25}Al_{0.25}La_3Zr_2O_{12}$ (C-LALZO) electrolyte via sol-gel process with graphene oxide (GO) as nucleation sites. The PEO-based composite electrolyte with C-LALZO as the active filler (PLC) presents excellent electrochemical performance. Garnet LALZO has high ionic conductivity at room temperature and good electrochemical stability against lithium metal.[26] Furthermore, the advantages of C-LALZO in enhancing the conductivity and inhibiting the growth of lithium dendrites in composite electrolytes were investigated. The assembled $LiFePO_4$/PLC/Li all-solid-state battery demonstrate excellent cycling stability and rate performance at both 50 ºC and 60 ºC. C-LALZO can further enhance the ionic conductivity of composite electrolyte due to the formation of continuous fast Li-ion transfer path, which can effectively promote the uniform deposition of lithium metal without the growth of lithium dendrites. The decomposition potential of composite electrolyte is widened to around 5.2V at 50 ºC versus metallic Li, which means the composite electrolyte could match higher voltage cathode material such as $LiNi_{1/3}Co_{1/3}Mn_{1/3}O_2$ (NCM) and $LiCoO_2$ (LCO). This work provides a facial sol-gel method to fabricate coral-like LALZO, and all-solid-state battery with its composite electrolyte achieves outstanding cycling stability and rate performance.

## Experimental Section

**Preparation of $Li_{6.25}Al_{0.25}La_3Zr_2O_{12}$ Branches and Corals:**

Al element was selected to stabilize cubic garnet phase. Stoichiometric amounts of $LiNO_3$, $La(NO_3)_3 \cdot 6H_2O$, $C_{20}H_{28}ZrO_8$, $Al(NO_3)_3 \cdot 9H_2O$ (purchased from Aladdin) were dissolved in a mixed solvent (ethanol: deionized water = 3:1, by volume), followed by adding citric acid (citric acid: metallic cation=2:1 in mole ratio) as the complexant. Additional $LiNO_3$ (20 wt %) was added to supply the lithium loss during the calcination. After that, the solution was stirred on a hot plate to evaporate solvent and formed a tawny solid precursor. To obtain the cubic phase branch-like LALZO (B-LALZO), the precursor was first heated at 200 °C for 2 h in vacuum to remove the residual solvent and then calcined at 800 °C for 5h in air. The preparation of coral-like LALZO (C-LALZO) was the same as that of B-LALZO except addition of GO ($LiNO_3$: GO=16 in weight) as the nucleation sites into the precursor solution.

**Preparation of Composite Electrolyte:**

PLC (PEO with C-LALZO) and PLB (PEO with B-LALZO) composite electrolytes were fabricated by solution casting method. PEO (Mw=600000, sigma) and LiTFSI (Aladdin, 99%) was added into acetonitrile with a mole ratio of EO: Li=8:1. After stirring for several hours, B-LALZO or C-LALZO was added into aforementioned solution and stirred for another several hours. The homogeneously suspended composite solution was poured into Teflon molds and dried in an argon filled glovebox. The obtained composite solid electrolyte membrane was further dried at 60 °C for 10 h in a vacuum oven to remove residual solvent, and then punched into certain sizes and shapes.

**Characterization of LALZO and Composite Electrolyte:**

X-ray diffraction (XRD) patterns of the B-LALZO, C-LALZO, and composite electrolytes were recorded with an X-ray diffractometer (Bruker D8 Advance, Germany) using Cu Kα radiation ($\lambda$ = 0.154 nm) in an angular range of 10 °–90 ° at a power of 40 kV. The morphology and interplanar spacing of LALZO materials were characterized by a scanning electron microscope (SEM, Phenom pro) and a high resolution transmission electron microscope (HRTEM, JEOL JEM-2100). Thermogravimetric analysis (TGA) was performed with a Mettler-Toledo TGA2 Thermo Analyzer from 30 to 900 °C at a ramping rate of 5 °C min$^{-1}$ under argon. The Fourier transform infrared (FTIR) spectra were performed on a BRUKER TENSOR 37.

**Electrochemical Characterization of Solid Composite Electrolyte:**

CR2032 cells were assembled by sandwiching the electrolyte between two stainless steel (SS) sheets to test the ionic conductivity of the solid composite electrolyte with different LALZO contents using electrochemical impedance spectroscopy (EIS) method. The measurements were carried out in a frequency range from 1 MHz to 100 mHz from 25 to 70 °C on Metrohm-Autolab B.V electrochemical station. The electrochemical stability of the composite solid electrolyte was investigated via linear sweep voltammetry (LSV) which was applied to a cell with a stainless-steel plate as working electrode and Li metal as the counter and reference electrode. The applied potential ranges from 2.5 to 6 V at a scanning rate of 0.1 mV s$^{-1}$ at 60 °C. The assembled Li/CPE/Li cells were galvanostatically cycled at current density of 0.02 and 0.05 mA cm$^{-2}$ on Land 2001A battery testing system for interfacial stability testing. And LiFePO$_4$/PLB/Li and LiFePO4/PLC/Li cells were assembled for the GITT measurement. The assembled all-solid-state cells were subjected to a sequence of charging at 0.1 C for 1 h followed by 0.5 h-rest on the land 2001A battery test system until the potential reached 4 V (vs Li/Li$^+$).

**Batteries Assembly and Electrochemical Characterization:**

The all-solid-state battery was assembled in 2032 coin cells, using PLC and PLB as electrolyte, LiFePO$_4$ (LFP) as cathode active material, and Li metal as anode. The cathode is composed of 80 wt% LFP, 10 wt% PVDF and 10 wt% super-P. The all-solid-state batteries were assembled in an argon-filled glove box. The cycle performance tests of all-solid-state lithium battery at 50 and 60 °C were carried out within a voltage range from 2.8 to 4.0 V (for LFP). The rate performances were acquired from 0.1 C to 1 C at 60 °C and 50 °C. All of the above electrochemical tests were carried out in the land 2001A battery test system.

## Results and Discussion

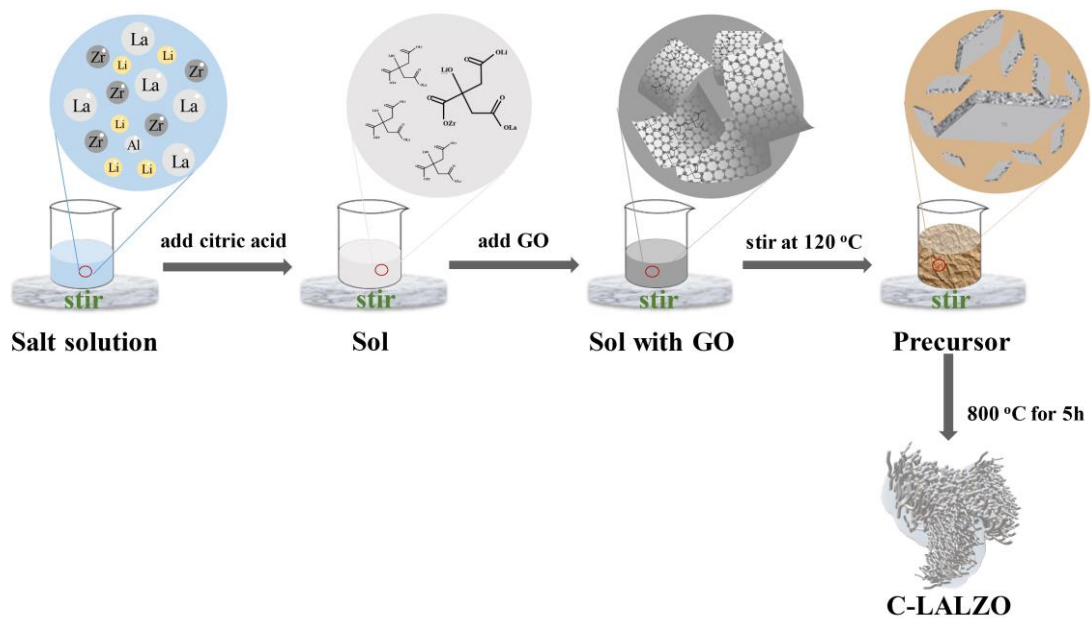

Figure 1. Schematic diagram of process to fabricating C-LALZO.

Sol-gel method to synthesize LALZO has been widely reported before. In this work, we first successfully developed coral-like LALZO with cubic garnet structure by using GO as nucleation site under sol-gel conditions, of which the mechanism was briefly discussed in the supporting information. The procedure of fabricating C-LALZO is illustrated as **Figure 1**.

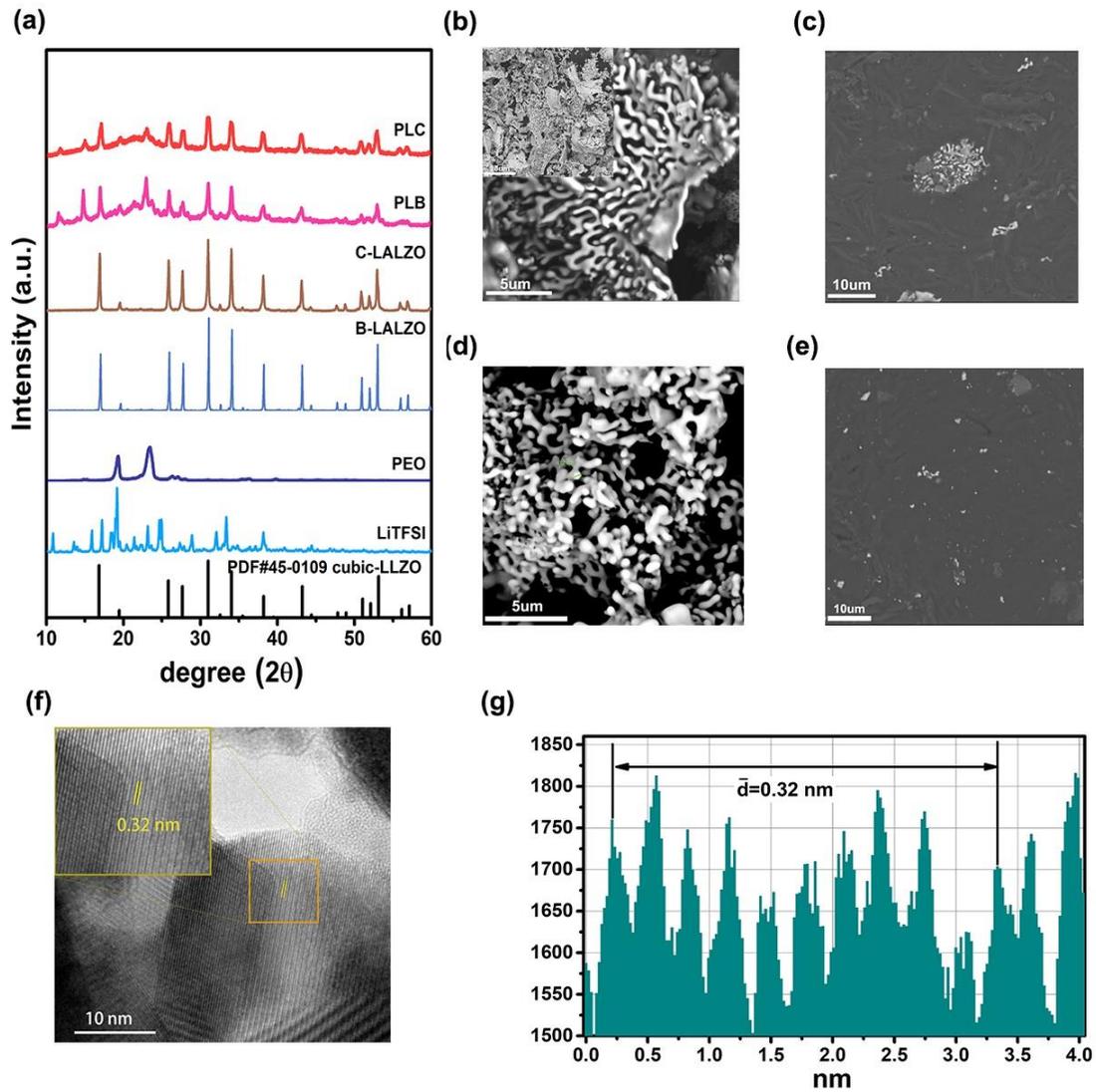

**Figure 2.** (a)XRD pattern of PEO, LiTFSI, LALZO and composite electrolytes with 20% C-LALZO and B-LALZO; SEM morphology of (b) C-LALZO (c)PLC (d) B-LALZO and (e) PLB; (g) HRTEM and (h) average interplanar spacing of LALZO coal.

The X-ray diffractions (XRD) in **Figure 2a** verifies the garnet crystalline structure of the C-LALZO and B-ALZO. Furthermore, as illustrated in **Figure 2f, g**, the HRTEM further reveals that the C-LALZO possesses a lattice space of 0.32 nm corresponding to the cubic phase LALZO (411) crystallographic plane, which further verifies the garnet crystalline structure of C-LALZO. Moreover, XRD spectrum of composite electrolytes with C-LALZO and B-LALZO are presented in **Figure 2a**. It can be observed that the peak intensity ratio of PEO in PLC is lower than that in PLB, which suggests C-LALZO provides more positive effect to reduce the crystallinity of PEO compared to B-LALZO. In addition, the XRD patterns of composite electrolytes with different C-LALZO contents are presented in **Figure S1**. With the increase of C-LALZO content, the intensity of peaks increases gradually. Besides, the peak intensity ratio of PEO is declined in composite electrolyte, which indicates that the C-LALZO filler effectively reduces the crystallinity of PEO. As presented in **Figure 2a**, the C-LALZO is in the form of 3D continuous coral, which serves as fast lithium ion transportation channel in the composite electrolyte to improve the ionic conductivity.[23] However, the LALZO branches are aggregated partly (**Figure 2d**). And it can be observed in **Figure 2c, e** that the garnet C-LALZO and B-LALZO particles are embedded in PEO electrolyte. Compared to B-LALZO, C-LALZO forms a one-dimensional (1D) and two-dimensional (2D) connected continue fast lithium ion conduction channel in PEO, which can promote the ion transportation of composite electrolyte.

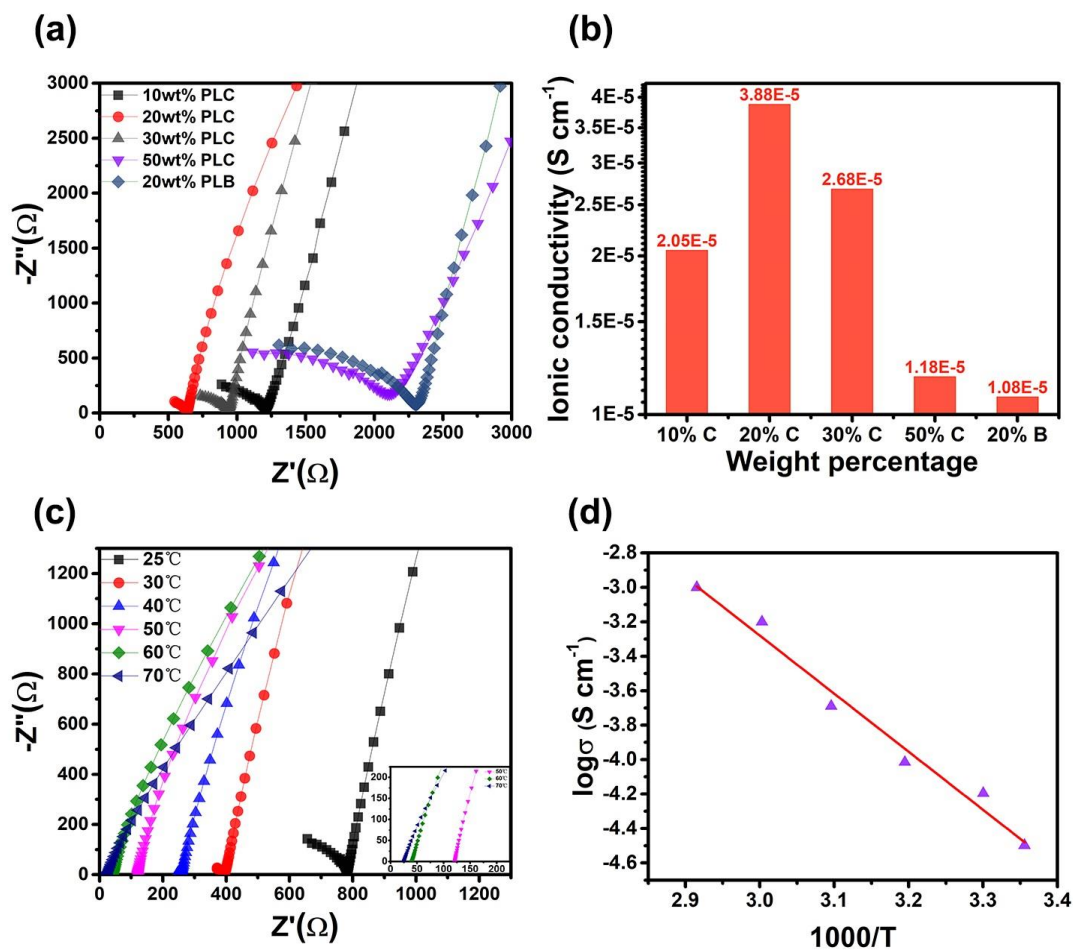

**Figure 3.** (a) EIS and (b) calculated ionic conductivity of composite electrolyte with different LALZO contents at 25 °C; (c)EIS and (d) Arrhenius plot of the PLC (20 wt% C-LALZO) at elevated temperatures (from 20 °C to 70 °C).

The ionic conductivity (σ) is calculated via the following equation:[27]

$$\sigma = \frac{L}{RS}$$

In which $S$ is the area of electrolyte, $L$ the thickness of electrolyte. $R$ is obtained by EIS measurement with symmetric cell of the electrolyte sandwiched by two stainless steel electrodes. Electrolytes with a thickness of ~400 μm are adopted for the ionic conductivity measurements to ensure precise thickness measurement. The Nyquist plots of PLC and PLB for evaluating the ionic conductivity are shown in **Figure 3a**.

According to the ionic conductivity of PLC electrolytes with various C-LALZO contents (**Figure 3b**), the PLC with 20 wt% C-LALZO has an optimum ionic conductivity. Therefore, for comparison, the ionic conductivity of PLB with 20 wt% B-LALZO is also listed in **Figure 3b**. The results in **Figure 3b** show that the garnet coral-like filler exhibits better ionic conductivity enhancement than their branch-like counterparts. Besides, the results of EIS also verify the conjecture from the test results of XRD in **Figure 2a**. The higher ionic conductivity is attributed to their unique coral like continuous structure with 1D LALZO nanowires seating on a 2D LALZO plate, which can promote the ion transportation and provide fast ion transportation routine. **Figure 3e** presents the ionic conductivity of PLC at different temperatures from 25 to 70 °C. The linear relationship **(Figure 3f)** between Li-ion conductivity and temperature conforms to the Arrhenius equation.[28]

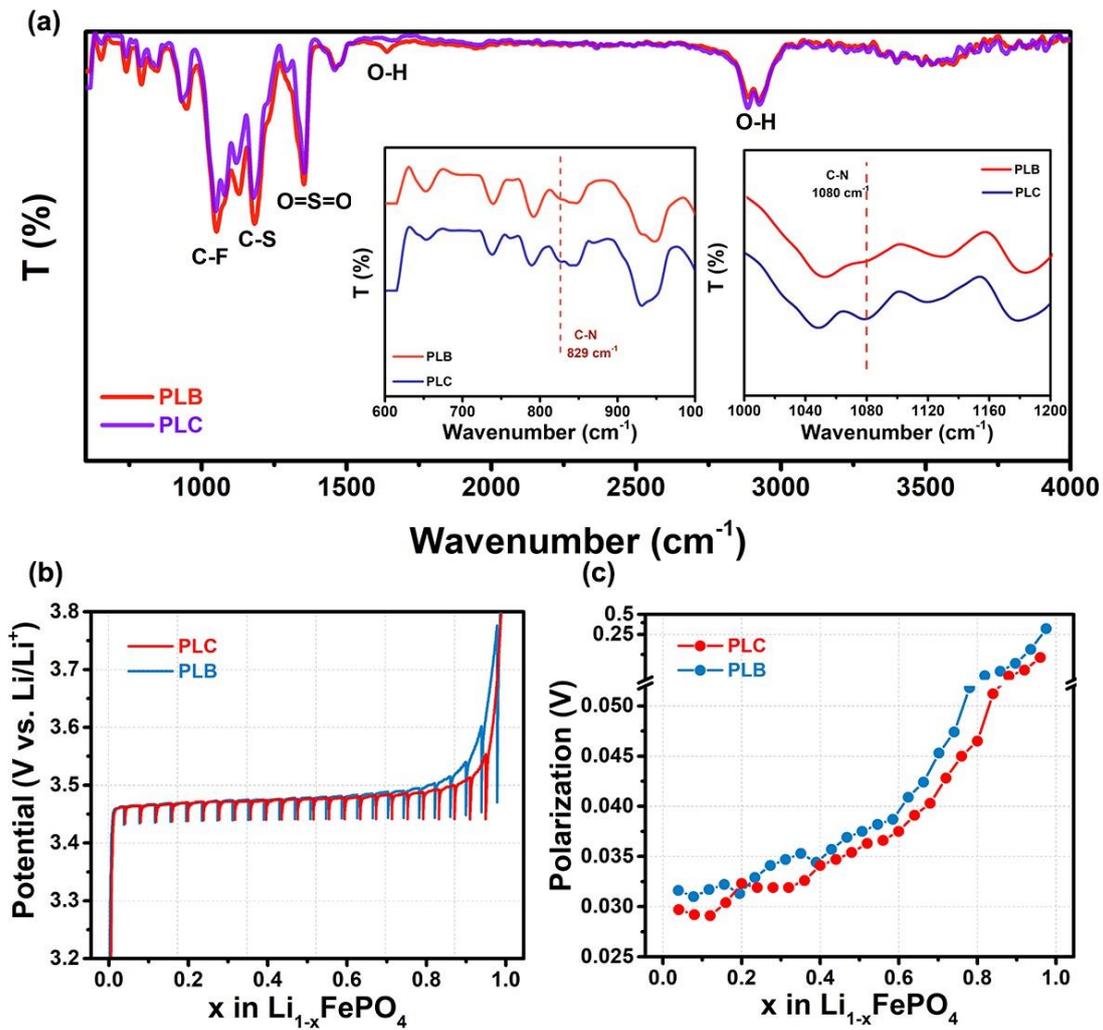

**Figure 4**. (a) FTIR of PLB and PLC (inset: detail of plot at around 829 and 1080 cm$^{-1}$). (b)GITT profiles and (c) the polarization of PLB and PLC at the different state of charging process

The FTIR spectrum in **Figure 4a** reveals that the enhanced peaks at 829 and 1080 cm$^{-1}$ appear in PLC composite electrolyte compared to that of PLB, which is related to the functional group of C–N.[29] The C-N group may be the result of the interaction between carbon and nitrogen containing functional groups. Because of the existence of ceramic LALZO, TFSI$^-$ anion could be fixed by PEO polymer chain, thus inducing the uniform distribution of space charge, which is beneficial to inhibit the growth of lithium dendrite.[30] GITT (Galvanostatic intermittent titration technique) experiments were performed at 60 °C to further solidify conclusion from the results of EIS and FTIR. The corresponding transient charge potential profiles and the calculated polarization are presented in **Figure 4b** and **c**, respectively. As shown in **Figure 4b**, the curve of battery with PLC electrolyte delivers smaller overpotential compared to that of PLB electrolyte, indicating a higher Li$^+$ diffusion coefficient (Ds). PLC electrolyte displays lower polarization values than that of PLB electrolyte at various potentials of the charging process **(Figure 4c)**, which suggests faster Li$^+$ diffusion kinetics in PLC electrolyte than that in PLB electrolyte.[31]

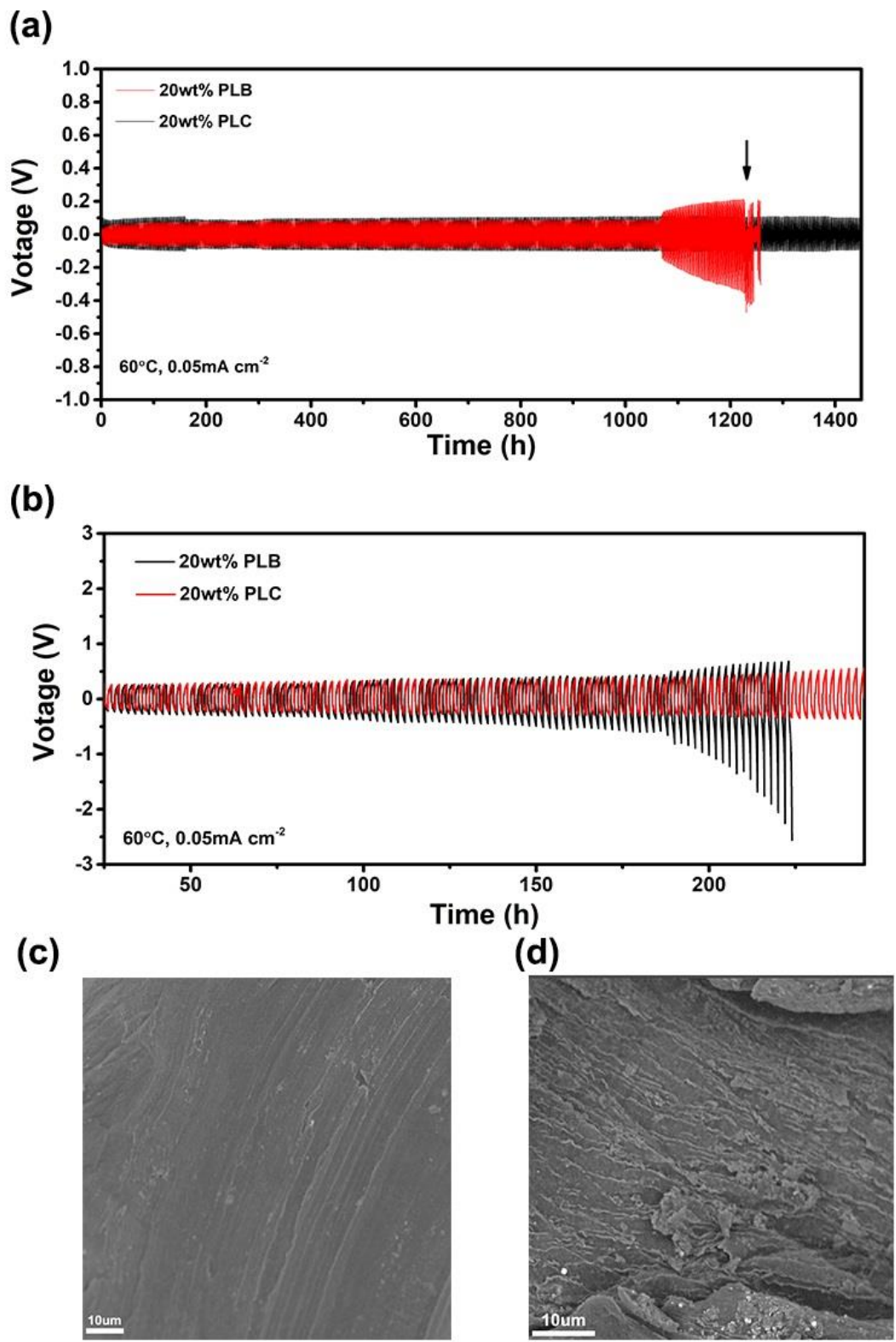

**Figure 5.** (a, b) voltage profiles of the lithium plating/stripping cycling with current density of 0.02 and 0.05 mA cm$^{-2}$ at 60 ℃. SEM of Li metal anode after cycle in (c) Li/PLC/Li and (d) Li/PLB/Li symmetric cells

The stability of the composite solid electrolyte against Li metal was evaluated by Li symmetric cells. The Li/PLC/Li and Li/PLB/Li symmetric cells were charged and discharged for 1 h under current densities of 0.02 and 0.05 mA cm$^{-2}$ at 60 ℃, respectively. (**Figure5a, b**). After cycling for 1500 h, the voltage hysteresis of symmetric cell with PLC can still be stabilized at about 100 mV. However, abrupt voltage drops (indicated by arrow in **Figure 5a**) are observed in the later cycling period of Li/PLB/Li due to the short circuit of the cell, which suggests undesired deposition of lithium. This result confirms that the PLC composite electrolyte is more effective to inhibit the lithium dendrite growth than the PLB electrolyte. Furthermore, when the current density is increased to 0.05 mA cm$^{-2}$, voltage polarization of Li/PLB/Li cells increased rapidly. This unfavorable cycling behavior is attributed to the inhomogeneous ionic conductivity and low ionic conductivity of PLB, which leads to the undesired lithium deposition. The polarization voltage of Li/PLC/Li cells keeps at 250 mV, suggesting good cycling stability with long cycle life and superior ability to suppress lithium dendrite growth. Besides, in order to further confirm the superior positive effect of PLC to inhibit lithium dendrite growth, the surface morphology of lithium anode obtained from symmetric cells after cycle for 1500h are presented in **Figure 5b, c**. It is obvious that the lithium anode in symmetric cells with PLC electrolyte exhibits smoother surface than that with PLB.

s

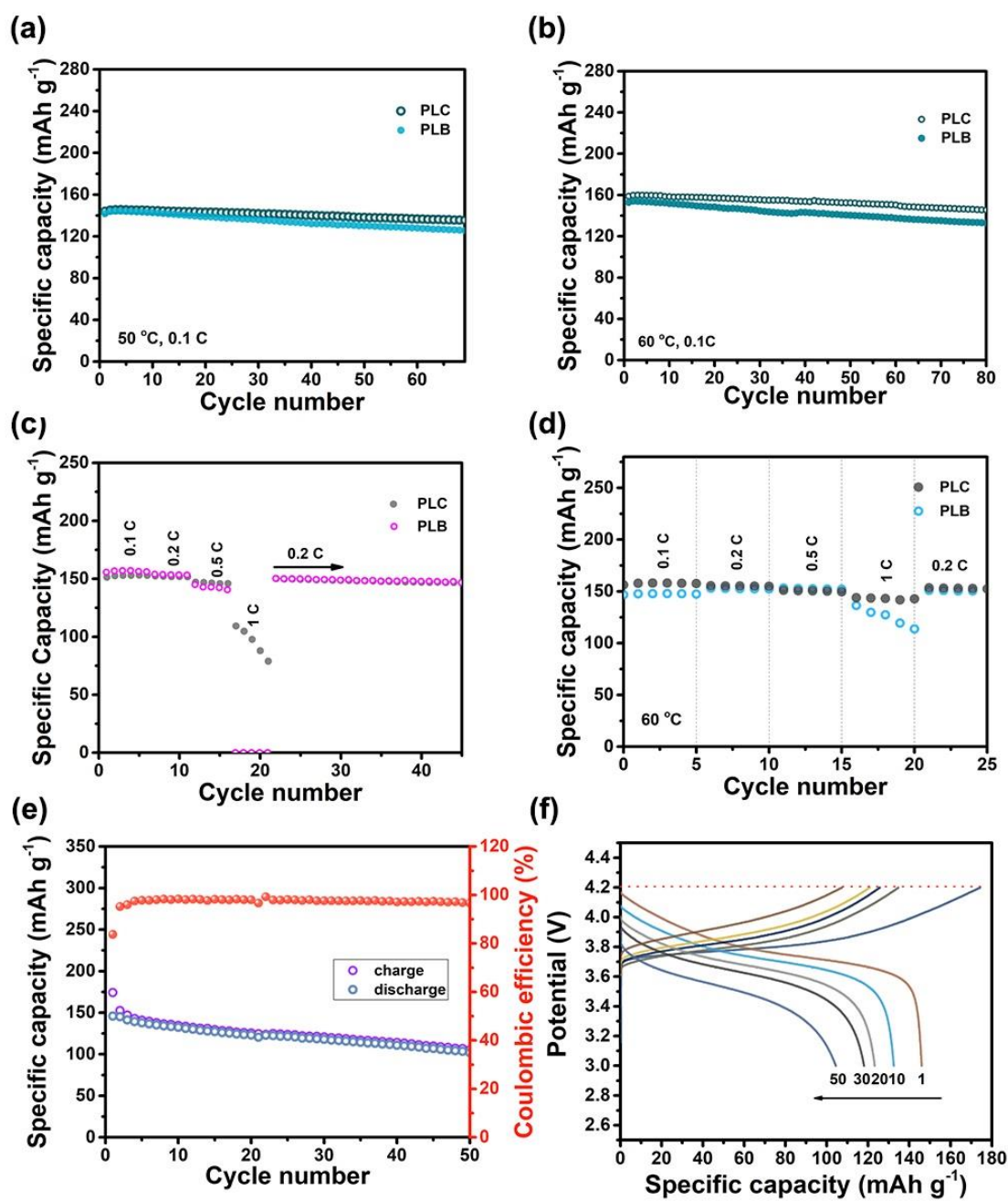

**Figure 6.** Electrochemical performances of LFP/PLC/Li and LFP/PLB/Li cells. Cycling performance of LFP/PLC/Li and LFP/PLB/Li cells at (a) 50 °C and (b) 60 °C. rate performance of LFP/PLC/Li and LFP/PLB/Li cells at (c) 50 °C and (d) 60 °C. (e)cycling performance and (f)charge–discharge voltage profiles of NCM111/PLC/Li at 0.1 C and 60 °C.

All-solid-state LFP/PLB/Li and LFP/PLC/Li batteries were assembled to verify the electrochemical advantages of coral like LALZO structure. **Figure 6a** and **b** present the cycling performance at 0.1 C at 50 °C and 60 °C, respectively. The assembled LFP/PLC/Li exhibits initial discharge specific capacity of 159.8 mA h g$^{-1}$, which remains at 145.6 mA h g$^{-1}$ after 80 cycles with a capacity retention rate of 91.1%. However, LFP/PLB/Li exhibits lower capacity retention rate of 86.3%. Since the operating temperature makes a great difference to the performance of all-solid-state batteries, the performance of all-solid-state batteries at a lower work temperature was measured. At 50 °C, the discharge capacity of LFP/PLC/Li cell exhibits 135.5 mAh g$^{-1}$ after 70 cycles with a capacity retention rate of 93.3% (**Figure 6a**), which is much higher than the LFP/PLB/Li cells with 125.6 mAh g$^{-1}$ after 70 cycles with a capacity retention rate of 86.5%. The significant promotion of specific capacity is attributed to the high conductivity of PLC ensuring fast ion transport and stable interfacial contact between cathode and PLC in the LFP/PLC/Li cells. The results suggest that the batteries with PLC can work at a lower temperature.

Rate performance of assembled all-solid-state batteries with PLC and PLB composite electrolyte at various rates from 0.1 C to 1 C at 50 °C and 60 °C is presented in **Figure 6c** and **Figure 6d**, respectively. The discharge specific capacity of LFP/PLC/Li cell cycled at 60 °C at 0.1 C, 0.2 C, 0.5C and 1.0 C is 158.7, 155.4, 150.5 and 143.9 mA h g$^{-1}$, respectively. (The discharge specific capacity of LFP/PLC/Li cell cycled at 50 °C at 0.1 C, 0.2 C, 0.5C and 1.0 C is 153.2, 150.8, 146.5 and 99.3 mA h g$^{-1}$, respectively) These specific capacity in different rates are higher than that of batteries with PLB composite electrolyte at corresponding rates, indicating fast ion transportation of PLC, which is consistent with the higher Li$^+$ diffusion coefficient and smaller overpotential shown in **Figure 6b**. The discharge specific capacity recovers to 153.7 mAh g$^{-1}$ at 0.2 C, 60 °C and 150.2 mAh g$^{-1}$ at 0.2C, 50 °C after cycling at 1C, suggesting excellent stability of all-solid-state LFP/PLC/Li batteries.

Furthermore, LSV (Linear Sweep Voltammetry) in **Figure S2** presents the decomposition voltage of PLC and PLB, which reveals the PLC possess wider electrochemical window than that of PLB. Therefore, high voltage cathodes (NCM and LCO) were adopted to assemble all-solid-state batteries. The all-solid-state batteries using NCM111 and LCO cathode exhibit excellent cycling stability at 0.1C and 60 °C. Besides, **Figure 6f** and **Figure S4** shows that the NCM/PLC/Li and LCO/PLC/Li cells present smooth charge–discharge curves, indicating that there are no side reactions occurred and PLC composite electrolyte possess ultrahigh electrochemical stability at high voltage.

## Conclusion

In summary, an all-solid-state battery is developed with composite electrolyte embedded with coral-like LALZO. The coral-like LALZO increases the ionic conductivity and enhances the electrochemical properties of PEO. This novel composite solid state electrolyte effectively suppresses the lithium dendrite growth and the Li symmetric cells can stably cycle about 1500 h without short circuit at 50 °C. The all-solid-state LFP/PLC/Li batteries present a specific capacity of 145.6 mAh g−1 after 80 cycles at 0.1 C under 60 °C and a specific capacity of 135.5 mAh g−1 after 70 cycles at 0.1 C and a lower temperature of 50 °C. This work provides a facial sol-gel method to fabricate coral-like LALZO, and develops an all-solid-state battery with its composite PEO electrolyte that achieves good cycling stability and rate performance. Furthermore, it provides a reference for researchers to study the synergistic electrochemical effects of active ceramic fillers of various dimensions and morphologies in PEO.

## Acknowledgements

This work was supported by High-level Talents' Discipline Construction Fund of Shandong University (31370089963078), Shandong Provincial Science and Technology Major Project (2016GGX104001, 2017CXGC1010, and 2018JMRH0211), the Fundamental Research Funds of Shandong University (2016JC005, 2017JC042 and 2017JC010), and the Natural Science Foundation of Shandong Province (ZR2017MEM002) materials for supercapacitors and sodium-ion batteries. *Journal of Power Sources* **2019**, *414*, 308-316.

# Supporting Information

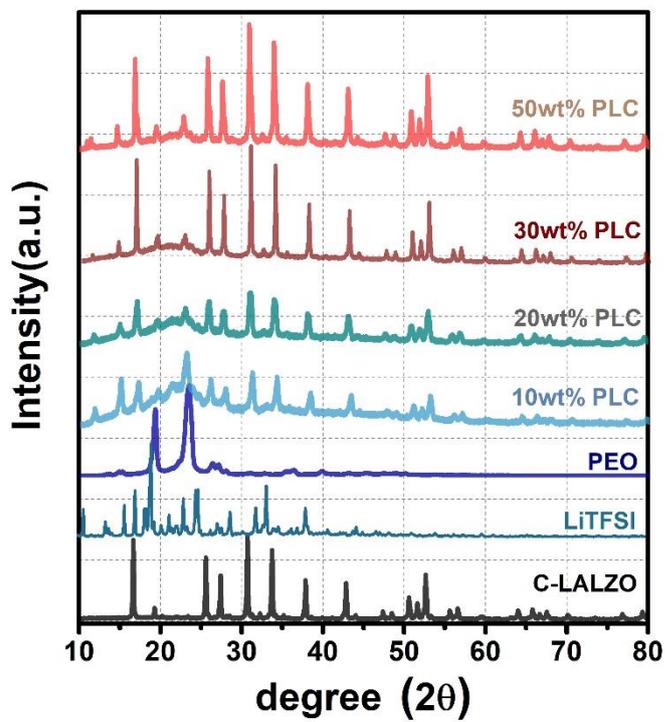

**Figure S1:** XRD of composite electrolytes with different coral LALZO contents

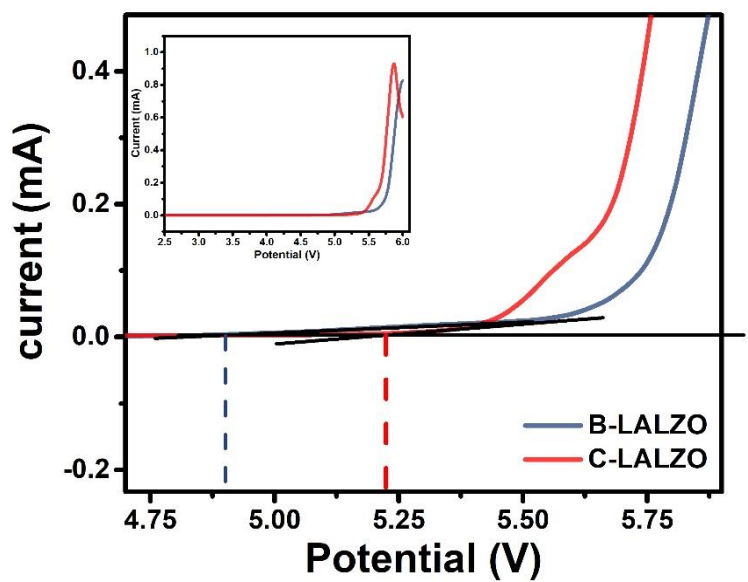

**Figure S2:** LSV of PLB and PLC

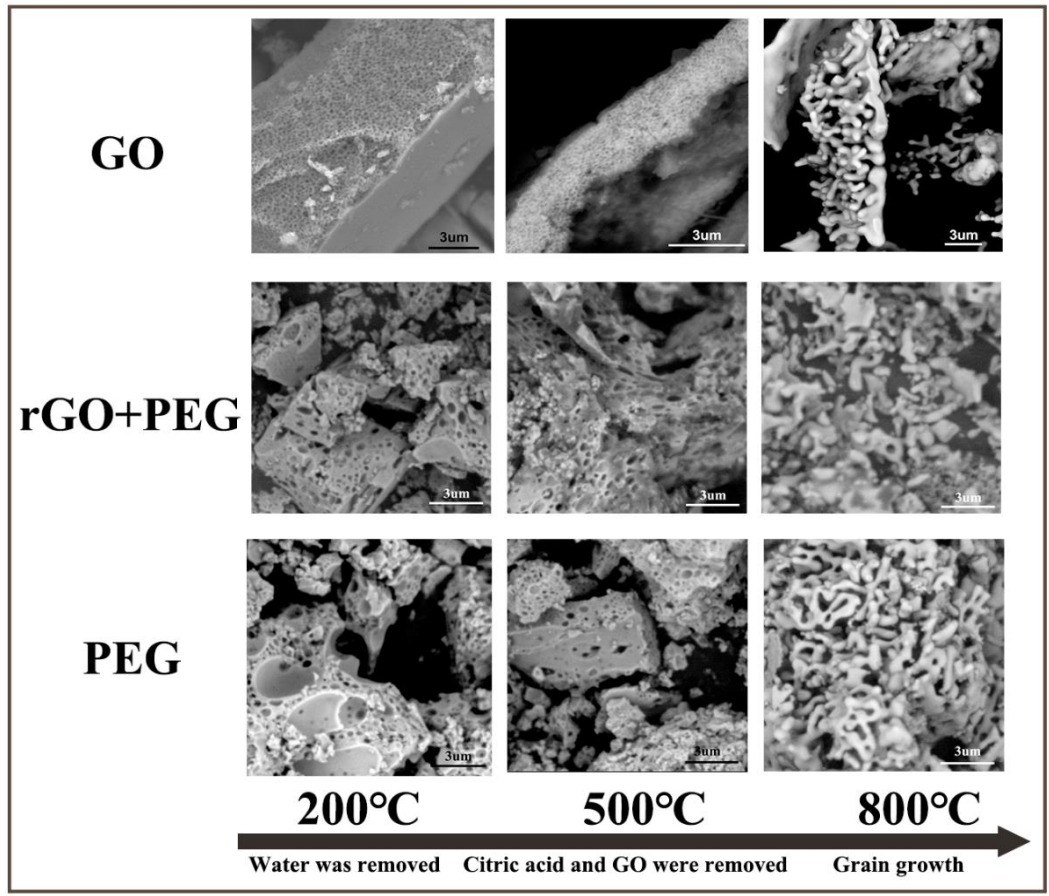

**Figure S3:** Morphology of LALZO with different nucleating agent at different stage of calcine

The effect of GO in fabricate C-LALZO can result in two points: 1. The high specific area of GO provides more active sites for metal cations to adhere; 2. The abundant functional groups on the surface of GO have a strong attraction to metallic cations. In order to investigate the formation mechanism of C-LALZO, the functional groups on the surface of GO was removed by hydrothermal method. To prevent GO-agglomeration after being reduced, excess PEG-2000 as dispersant was added. Furthermore, another contrast group with pure PEG-2000 was developed to eliminate the effect of PEG. As presented in **Figure S3**, SEM of C-LALZO shows when residual water was removed at 200 °C, compare to the other two samples, the precursor of sample with GO exhibits a honeycomb-like morphology while there are no obvious honeycomb-like structure in the sample with rGO+ PEG or PEG. In addition, there are smooth plant on one side of honeycomb-like precursor in sample with GO. After the temperature increased to 500 °C, the precursors are gradually shrinking due to the decomposition of citric acid and GO. At 800 °C, the SEM presents C-LALZO have been developed in sample with GO. Whereas, there no obvious coral like LALZO structure can be found in the other two samples. The above results suggest the formation of C-LALZO result from the strong attractive force from abundant functional groups and metallic cations.

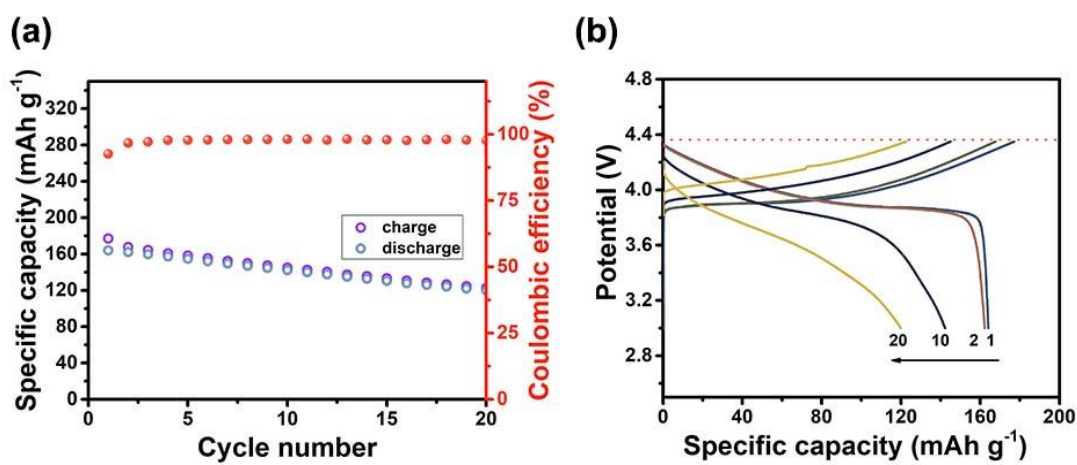

**Figure S4.** (a) Cycling performance and (b) charge–discharge voltage profiles at 0.1 C and 60 °C of LCO/PLC/Li